\documentclass[12pt]{article}
\usepackage{amsmath,graphicx}

\def\eq#1{Eq.~\ref{#1}}
\def\fig#1{Fig.~\ref{#1}}

\begin{document}

\title{Collective effects in cellular structure formation mediated by compliant environments:
a Monte Carlo study}
\author{Ilka B. Bischofs\footnote{Present address: Department of Bioengineering,
Center for Synthetic Biology, University of California,
Berkeley, CA 94720-3224, USA} $$ and Ulrich S. Schwarz
\\[1cm]
University of Heidelberg, Im Neuenheimer Feld 293,\\
D-69120 Heidelberg, Germany\\[1cm]
Max Planck Institute of Colloids and Interfaces,\\
D-14424 Potsdam, Germany}

\date{}
\maketitle
\thispagestyle{empty}


\newpage
\thispagestyle{empty}

\begin{abstract}
  Compliant environments can mediate interactions between mechanically
  active cells like fibroblasts. Starting with a phenomenological
  model for the behaviour of single cells, we use extensive Monte
  Carlo simulations to predict non-trivial structure formation for
  cell communities on soft elastic substrates as a function of elastic
  moduli, cell density, noise and cell position geometry.  In general,
  we find a disordered structure as well as ordered string-like and
  ring-like structures. The transition between ordered and disordered
  structures is controlled both by cell density and noise level, while
  the transition between string- and ring-like ordered structures is
  controlled by the Poisson ratio. Similar effects are observed in
  three dimensions. Our results suggest that in regard to elastic
  effects, healthy connective tissue usually is in a macroscopically
  disordered state, but can be switched to a macroscopically ordered
  state by appropriate parameter variations, in a way that is
  reminiscent of wound contraction or diseased states like
  contracture.
\end{abstract}

\newpage
\setcounter{page}{1}

\section{Introduction}

In order to develop and maintain tissues, cells in multicellular
organisms have to interact with each other and the extracellular
matrix (ECM). Cellular communication proceeds mainly through specific
interactions provided by receptor-ligand binding. In solution,
gradients in ligand concentration encode spatial information. For
example, morphogen gradients guide cell differentiation
\cite{c:turi52}, and gradients of chemoattractants or chemorepellants
direct cell motility (\textit{chemotaxis}) \cite{c:berg77}. For cells
adhering to each other or to the ECM, physical
properties of the environment like topography and mechanics provide
additional information supplementing specific biochemical cues.  In
particular, cells in their natural environment typically orient along
fiber bundles of the ECM, a principle termed \textit{contact guidance}
\cite{Weiss34}. In general, cells preferentially orient along
directions with minimal curvature \cite{Dunn76}. While contact
guidance provides only a bidirectional cue for cell migration,
unidirectional cues result from spatial gradients in adhesiveness
(\textit{haptotaxis}) \cite{Carter67}, substrate rigidity
(\textit{durotaxis}) \cite{Lo00} and substrate tension
(\textit{tensotaxis}) \cite{Beloussov00,Mandeville97}.

Tissue cells like fibroblasts in the connective tissue constantly
remodel their structural environment by degrading old and secreting
new ECM. Moreover they can exert forces large enough to actively
reorganize the ECM after it has been laid down \cite{Harris80}. Hence,
cells not only respond to the physical properties of their
environment, but also actively modulate them.  This results in
indirect, matrix-mediated interactions between cells. In particular,
cell traction-induced reorganization of collagen fibers can mediate a
mechanical interaction between cells via contact guidance
\cite{Murray83,Murray84}. In the same vein, elastic interactions
between cells can result from stress and strain in the ECM induced by
cell traction. During recent years, the sophisticated use of elastic
substrates has proven that cells indeed respond to purely
elastic features in their environment, including rigidity, rigidity
gradients and prestrain
\cite{Pelham97,Lo00,Wong03,Engler04,Engler04b,c:yeun05}.  It now
appears that many cell types, including fibroblasts, smooth muscle
cells, and endothelial cells (but not neutrophils or neurons), respond
to the mechanical properties of their environment with a common
preference for large effective stiffness
\cite{Wong04,c:geor05,c:disc05}. Here the term \textit{effective
  stiffness} comprises both rigidity of and tensile strain in the
environment, which can be actively sensed by cells via
mechanotransductory processes at cell-matrix adhesions
\cite{Balaban01,Riveline01,Geiger02,Bershadsky03}.  These mechanical
cues play an important role in a variety of physiological processes,
including development, tissue maintenance, angiogenesis, myotube
fusion, wound healing and metastasis
\cite{Galbraith98,Huang99,Geiger01a,c:bao03}. In particular, if
coupled to cell division, stress and strain become major determinants
of tissue morphogenesis \cite{c:shra05,c:nels05}. In general,
tissue function arises from the close relationship between
cell behaviour in and material properties of the ECM
\cite{Curtis01,Griffith02,Langer04}.

We have argued before that elastic interactions contribute to the way
single cells position and orient themselves in compliant environments
\cite{Schwarz02a,Bischofs03,Bischofs04}.  However, it has not been
discussed in detail before how this individual behaviour translates
into collective behaviour of cell ensembles in compliant environments.
In a recent short report we have shown that elastically interacting
cells assemble into strings of cells which then leads to screening of
the cellular traction patterns \cite{uss:bisc05a}.  Although this
effect in principle suppresses macroscopic order, we have also found
that macroscopic order can exist at high cell density, with an
interesting competition between string- and ring-like structures as a
function of the Poisson ratio of the surrounding material. In our
recent report, this effect has been especially analyzed for ordered
(lattice) arrangements of cells, which can be decomposed into strings.
Since cellular systems are intrinsically noisy, in this paper we focus
on stochastic effects using extensive Monte Carlo simulations, which
allow to predict collective effects in the presence of noise. In the
following, we mainly consider the case of cells adhering to the top of
a soft elastic substrate, because this setup might be the easiest way
to experimentally verify our theoretical predictions.  Commonly used
materials for elastic substrates are polyacrylamide or
polydimethylsiloxane, which can be described by linear isotropic
elasticity theory.  Since we are interested in the interactions
mediated by an elastic environment, we only consider situations
without cell-cell contact, which experimentally could trigger
different cellular responses to mechanical signals.  Therefore in our
theoretical work we first fix cell positions and then relax cellular
orientations using standard Monte Carlo techniques.  Experimentally,
this might be done by using microcontact printing, thus restricting
cells to adhesive islands, or by using non-motile cell lines, which
adhere at random positions and then do not move.  By freezing in cell
positions, we are also able to control cell density.  For the position
geometry, we consider ordered (lattice) arrangements, random
perturbations around ordered positions and arrangements which are
completely random. We calculate structural phase diagrams as a
function of of elastic moduli, cell density, noise and cell position
geometry using Monte Carlo simulations. We then briefly investigate
collective elastic effects in three dimensional elastic environments
and finally discuss our results in the context of wound healing.

\section{Model and simulations}

\subsection{General concepts}
In order to sense the mechanical properties of their environment,
tissue cells pull on it with actomyosin contractility.  In many cases,
this contractile mechanical activity is directed along the long axis
of the cell body, especially when confronted with a mechanically
anisotropic environment. The mechanical action of such an anisotropic
force pattern can be modeled as an anisotropic force contraction
dipole $P_{ij}= P l_i l_j$, where the unit vector $\vec l$ specifies
cell orientation. In contrast to electric dipoles in electrostatics,
which are vectors, force dipoles in elasticity theory are tensors of
rank two. $P$ specifies the dipole strength, which typically has $|P|
= F d \approx 10^{-11}$ J, corresponding to two opposing forces $F =
200$ nN separated by a distance $d = 60$ $\mu$m \cite{Schwarz02b}. We
will consider cells with anisotropic force patterns only and assume
that the magnitude $P$ is constant for all cells. A mechanically
active cell generates an elastic strain field $u_{ij}(\vec r)$, which
again is a tensor of rank two. The strain induced by $P$ can be
calculated from the elastic Green tensor
$G_{ij}(\vec{r},\vec{r}^{\prime})$ which describes the deformation of
the medium at $\vec{r}$ caused by a point force at $\vec{r}^{\prime}$.
In general $G_{ij}$ depends on material properties and boundary
conditions. For translationally invariant situations, one has
$G_{ij}(\vec{r},\vec{r}^{\prime})=G_{ij}(\vec{r}-\vec{r}^{\prime})$.
Elastic interactions between two cells result if the mechanical
activity $P_{ij}$ of one cell responds to the elastic field $u_{ij}$
induced by another cell. In the framework of a general field theory
elastic interaction can then be represented by a coupling of $u_{ij}$
and $P_{ij}$.  Since cells are active and moreover often show a
regulated response, the exact form of the coupling between $P_{ij}$
and $u_{ij}$ cannot be predicted from first principles. In a
first order approximation we may assume that $P_{ij}$ and $u_{ij}$ are
linearly coupled.  For mechanically active cells like fibroblasts,
experimental observations suggest that indeed they adopt positions and
orientations in such a way as to effectively minimize the scalar
quantity $W = P_{ij}u_{ij}$ \cite{Bischofs03,Bischofs04}.  For tensile
strain, this implies that contractile cells actively align with the
external field.  Because they pull against the external stretch, cells
reduce displacement.  For fibroblast-like cells, this behaviour might
have evolved in the context of wound healing, when cell traction is
required to close wounds. A biophysical interpretation of the origin
of the extremum principle regarding $W$ is to note that $W$ has
dimensions of energy. Physically $W$ can be interpreted as an extra
energy contribution to the deformation energy required to build up a
force dipole in the presence of strain $u_{ij}$
\cite{Bischofs03,Bischofs04}.  Using the analogy of a harmonic spring
the basic idea can be explained easily: for a given spring constant
$K$, it takes the work $W = F^2 / (2 K)$ to build up the force $F$.
Therefore larger stiffness $K$ corresponds to smaller work $W$
required to reach the force $F$. In this way $W$ may be interpreted as
the inverse of an effective stiffness of the environment and the
extremum principle in $W$ corresponds to the experimental observation
that cell-matrix contacts grow stronger in a stiff environment, thus
eventually determining cell orientation. Using this extremum
principle, we have been able before to unify many diverse observations
which have been reported for the organization of single cells on
elastic substrates and in physiological hydrogels
\cite{Bischofs03,Bischofs04}. For example, it predicts that single
cells prefer to align in parallel and perpendicular to free and
clamped surfaces of finite sized samples, respectively.  It is however
important to note that the cellular response to the actively sensed
effective stiffness of the environment is very different from the
cellular response to cyclic external stretch with a 1 Hz frequency,
which is relevant for the physiology of lung tissue and the
cardiovascular system.  In this case, many cell types tend to orient
away from the direction of stretch \cite{c:shir89,c:wang00}, possibly
to avoid the recurrent deformation of their cytoskeleton.

For two cells, the above reasoning leads to the following
effective interaction potential \cite{Bischofs03,Bischofs04}
\begin{equation} \label{eq:two_particle_interaction}
W = P_{ij} u_{ij} = - \sum_{i,j,k,l} P_{ij} 
\frac{\partial}{\partial x_j}\frac{\partial}{\partial x_l}
G_{ik}(\vec{r}-\vec{r}^{\prime}) P^{\prime}_{kl}
\end{equation}
where we consider only situations with translational invariance.
The optimal cell configuration is described by the minimum of $W$ in
regard to cell positioning and orientation. Since in general $G_{ik}$
scales as $\sim 1/(E r)$, where $r$ is distance and $E$ an elastic
modulus, $W$ scales as $\sim P^2/(E r^3)$ and has units of energy.
Since in a linear material the cellular strain fields superimpose, the
functional that describes elastic interactions of a system of $N$
cells reads:
\begin{equation}
\label{eq:1Dstacka}  W =\frac{1}{2N}\sum_{\gamma = 1}^N
\sum_{\delta \neq \gamma}^N W_{\gamma\delta},
\end{equation}
where $W_{\gamma\delta}$ is the interaction between two cells $\gamma$
and $\delta$ as described in \eq{eq:two_particle_interaction}. The
factor $1/2$ is required to avoid double counting and the factor $1/N$
for \textit{per cell} normalization. The optimal structure is given by
the configuration which minimizes \eq{eq:1Dstacka} as a function of
all cellular orientations and positions and we may refer to this
structure as the \textit{ground state}, in analogy to interacting
passive particles in physical systems.

\subsection{Monte Carlo simulations}

In the presence of noise any system will deviate from its optimal
state. In the cellular systems discussed here, noise results both from
intracellular processes (like the intrinsic stochasticity of gene
expression and signal transduction) and from heterogeneities in the
material properties of the environment.  In order to introduce a
stochastic element into the structure formation process, we perform
Monte Carlo simulations using the standard Metropolis algorithm to
generate typical configurations in the presence of noise
\cite{AllenMP01}. This implies the use of an Gibbs ensemble, which in
our case is the simplest choice given that we do not know the exact
details of the cellular decision making process (in information
theory, the Gibbs ensemble arises from maximizing the measure for
disorder (\textit{Shannon entropy}) under the constraint of a
prescribed average energy). Starting from an arbitrary configuration,
a cell is selected at random and its orientation randomly varied. The
new configuration is always accepted when it decreases $W$. Otherwise
it is accepted with the probability $\exp \left( - \Delta W / k_B T
\right)$, where $k_B$ is the Boltzmann constant and $T$ is
temperature.  In our context $k_B T$ represents a measure for the
degree of stochasticity involved in cellular decision making.  Other
studies on modelling cellular structure formation have used the same
concept before and found an effective value of $k_B T = 5\ 10^{-15}$ J
\cite{Drasdo95,c:dras00,Beysens00}. Since this quantity comprises all
molecular interactions between a cell and its environment, it is six
orders of magnitudes larger than thermal energy. In our case, the
competition between elastic and stochastic effects can be
characterized by the reduced temperature
\begin{equation} \label{eq:Tstar}
T^{\star}=\frac{k_B T \pi E b^3}{P^2}
\end{equation}
which measures the relative importance of noise with respect to the
average elastic interaction strength. Here $b$ is the average distance
between two cells which is related to the averaged cell density by
$\langle \rho \rangle = 1/b^2$ in two dimensions (2D) and by $\langle
\rho \rangle =1/b^3$ in three dimensions (3D). As we will see below,
typical values in our simulations range from $T^{\star} = 0.1$ to
$3.0$. Using $P = 10^{-11}$ J, $E = 10$ kPa, $b = 100\ \mu$m and a
typical value $T^{\star} = 1$, we obtain $k_B T = 3\ 10^{-15}$ J.
Therefore our choice for the noise level is exactly in the same range
as the earlier estimates \cite{Drasdo95,c:dras00,Beysens00}. As the
effective temperature is decreased towards cero,
$T^{\star}\rightarrow0$, $W$ decreases and the ground states become
increasingly favorable.  This process is known as \textit{simulated
  annealing} and is often used to identify optimal structures
\cite{Kirckpatrick83} .  In the vicinity of an optimal structure,
almost all Monte Carlo moves are rejected.  In contrast, when
$T^{\star}\rightarrow\infty$, every Monte Carlo move is accepted and
the structures get disordered.  One Monte Carlo sweep corresponds to
$N$ such Monte Carlo moves. After thermal equilibration (typically
about $10^3-10^4$ Monte Carlo sweeps), the Metropolis algorithm
samples the important configurations typical for a given temperature
$T^{\star}$.  One can then calculate statistical properties of
structures at defined noise level by averaging the quantity of
interest over many configurations generated by the Metropolis
algorithm.

To study stochastic effects using our Monte Carlo simulations we
typically consider $N \approx 1000$ dipoles. In order to minimize the
effects of boundaries and finite size, we apply periodic boundary
conditions (pbcs), such that each dipole has the same number of next
neighbors and experiences the same local geometry. We implement pbcs
using the minimal image convention, i.e.\ we only consider the
interactions of the dipole with its $N-1$ nearest (image) dipoles
\cite{AllenMP01}. Note that in the two-dimensional situation of cells
on top of an elastic half space, the $1/r^3$ interaction effectively
constitutes a \textit{short-ranged} potential, because its area
integral does not diverge with system size ($\int^{L} r dr (1/r^3)
\sim 1/L$ where $L$ is system size). Thus, boundary effects are
expected to play only a minor role in this case and the minimal image
convention is a good approximation.

\subsection{Additional assumptions}

We now make some additional assumptions which will simplify our
subsequent calculations and which will allow for a direct comparison
of our theoretical calculations with appropriate experiments. First,
we assume that the environment can be described by isotropic linear
elasticity theory, which is a reasonable assumption for the synthetic
substrates like the ones made from polyacrylamide or
polydimethylsiloxane commonly used to study mechanical effects in cell
culture \cite{c:demb99,Schwarz02b}. Thus there are two elastic
constants: the Young modulus $E$ describes the rigidity of the
material and the Poisson ratio $\nu$ the relative importance of
compression and shear.  In particular we mainly consider the situation
of cells adhering to the top surface of a thick elastic film.  Such a
situation can be theoretically represented by an elastic half-space
geometry with dipolar orientations constrained to the x-y-plane.
Therefore the Young modulus $E$ and the Poisson ratio $\nu$ are 3D
quantities. The maximal value for $\nu$ is $1/2$, e.g.\ for strongly
hydrated polymer gels.  If such a material is tensed in one direction,
the shear mode is excited and it contracts in the perpendicular
directions (\textit{Poisson effect}). For common materials, the
minimal value for the Poisson ratio is $\nu = 0$, e.g.  in dehydrated
fibrous polymer gels.  Then the volume mode prevails and uniaxial
tension does not translate into lateral contraction. For cells
exerting tangential forces on top of an elastic substrate one can use
the Boussinesq Green function of an elastic halfspace \cite{b:land70}
to specify \eq{eq:two_particle_interaction} to \cite{Bischofs04}:
\begin{equation}
W_{\gamma\delta}= \frac{a_1 P^2}{r^3}
f(\theta,\theta^{\prime},\alpha) \label{2Dinteraction}
\end{equation}
where $r$ is the distance between cells and the cellular orientations
$\theta$, $\theta^{\prime}$ and $\alpha$ are defined via the scalar
products $\cos \theta = \vec l \cdot \vec{r}$, $\cos
\theta^{\prime}=\vec l^{\prime} \cdot \vec{r}$ and $\cos \alpha =\vec
l \cdot \vec l^{\prime}$. $f$ is given by
\begin{eqnarray}
f(\theta,\theta^{\prime},\alpha)&=&
3(\cos^2\theta+\cos^2\theta^{\prime}-5\cos^2\theta\cos^2\theta^{\prime}
-\frac{1}{3}) \nonumber \\
&-&(1-a_2)\cos^2\alpha -3(a_2-3)\cos\alpha\cos\theta\cos\theta^{\prime}
\label{2Dangle}
\end{eqnarray}
where $a_1 =\nu (1+\nu)/(\pi E)$ and $a_2 =(1-\nu)/\nu$. 

Secondly, because here we focus on elastic effects, we want to avoid
cell-cell contacts, which are known to change the mechanical state of
adhering cells \cite{c:yeun05}. Therefore we attribute an exclusion
disc of radius $a$ to each cell. Moreover in most of our simulations
cell positions are frozen and only orientational degrees of freedom
are considered. Experimentally, this situation might be achieved by
using non-motile cell lines or appropriate drugs to suppress cell
motility. Alternatively, one might use microcontact printing to
prepare well-defined adhesive islands constraining cell positions.

Finally, while the focus of our paper is on structure formation on
planar synthetic substrates to allow a direct comparison of theory and
experiment, we also briefly consider cells in a three-dimensional (3D)
environment. For computational simplicity we keep the assumption of
linear isotropic. For cells embedded in an infinite 3D isotropic
elastic environment, the interaction law stays the same as
\eq{2Dangle}, but with different constants $a_1=(1+\nu)/(8 \pi E
(1-\nu))$ and $a_2=(3-4 \nu)$ which now follow from the Kelvin
solution for the full elastic space \cite{b:land70}. In 3D, the
elastic interactions are truly long-ranged, in the sense that now the
volume integral over the elastic interaction diverges with system size
($\int^{L} r^2 dr (1/r^3) \sim \ln L$ where $L$ is system size), thus
boundary effects become more important.

\section{Results and Discussion}

\subsection{Structure formation for dipolar lattices}

\subsubsection{Phase diagrams under thermal noise}

We first consider structure formation in the presence of noise when
cells adopt regular positions on an infinite square (s) and hexagonal
(h) lattice.  In general we find a strong dependence of structure
formation on lattice geometry, Poisson ratio $\nu$ and noise level
$T^{\star}$. In particular our simulations show an interesting competition
between \textit{string-like} and \textit{ring-like} structures at low
noise levels, which we have also found before in a detailed analysis
of optimal structures \cite{uss:bisc05a}.  For the square lattice low
noise structures are always string-like. The lattice geometry favors
the formation of strings directed along one of the principal lattice
vectors $\vec n=(1,0)$ or $\vec n=(0,1)$ for all $\nu$. The limit $\nu
\rightarrow 0$ shows structural degeneracy, that is, additional
structures, e.g.\ parallel strings aligned along $\vec n=(1,1)$,
become equally favored as parallel (1,0)-string structure.  Enhanced
structural degeneracy of string-like structures in the limit of
vanishing Poisson ratio occurs also on the hexagonal lattice. Most
interestingly, however, on the hexagonal lattice string- and ring-like
structures exchange stability as a function of Poisson ratio. For $\nu
> 0.32$ the Poisson effect favors ring-like structures over strings,
with rings composed of four dipoles each rotated by 90 degrees
relatively to its neighbor. On both lattice types the orientational
order imposed by elastic interactions is gradually destroyed with
increasing noise levels. When elastic interactions dominate, dipoles
typically fluctuate weakly around their optimal orientations forming
long-ranged ordered structures. At intermediate noise levels,
long-ranged order is decreased, but there are still ordered domains
which locally adjust to elastic signals. As noise increased further,
domain size decreases and local fluctuations become stronger,
eventually giving eventually rise to completely random dipole
arrangements.
 
\begin{figure}
\begin{center}
\includegraphics[width=\columnwidth]{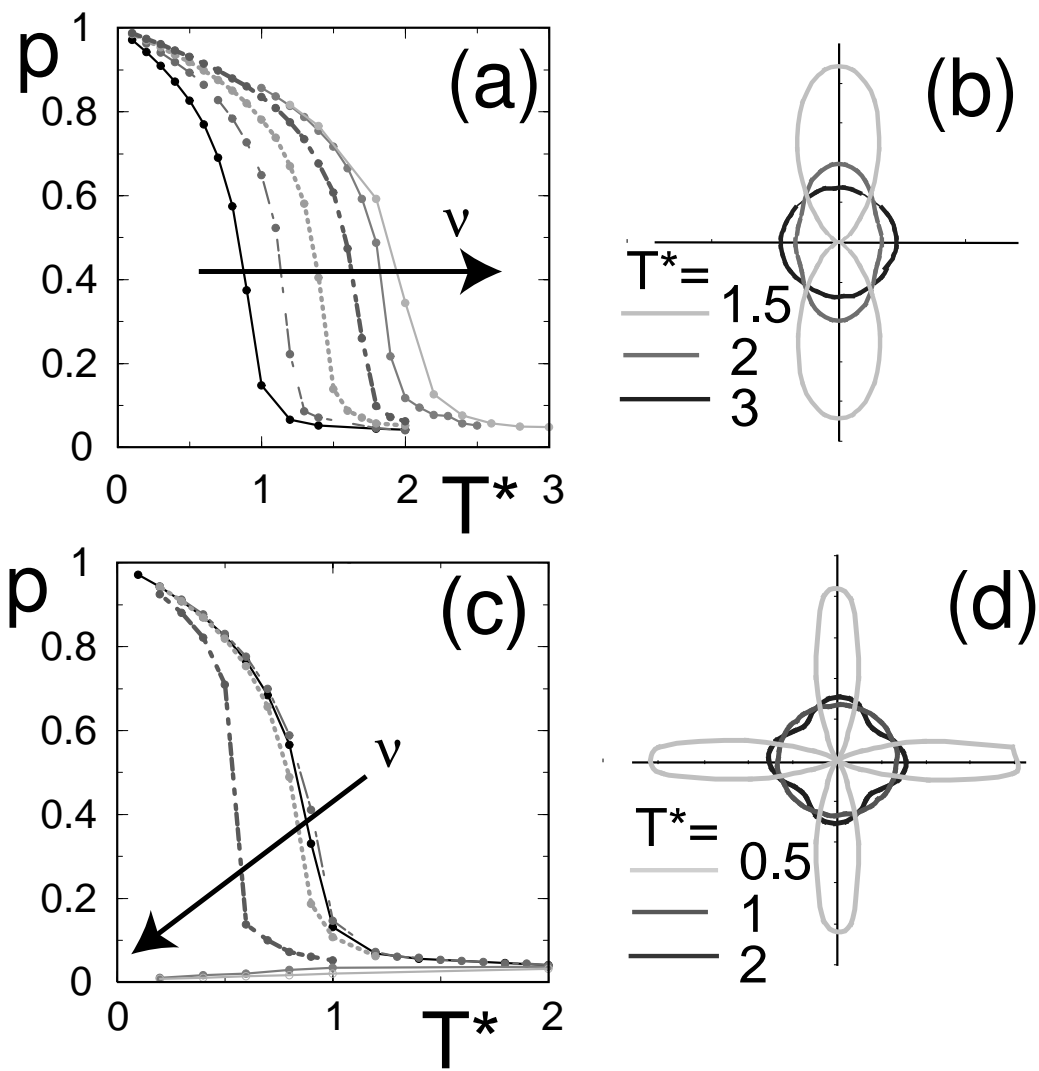}
\end{center}
\caption{Bischofs and Schwarz}
\label{fig1}
\end{figure}

The loss of long-ranged order with increased noise can be quantified
by calculating the temperature behavior of appropriate order
parameters. For string-like structures all dipoles point along a
common direction $\vec n$.  A suitable parameter to quantify the
degree of global alignment therefore seems to be the angle $\beta$
between dipole orientation $\vec l$ and $\vec n$, where $\cos \beta =
\vec l \cdot \vec n$. Note that force dipoles have a bipolar symmetry
and thus dipole orientations $\vec l$ and $-\vec l$ are equivalent.
Thus, the average $\langle \cos \beta \rangle=0$. However, the average
$\langle \cos^2 \beta \rangle$ becomes one when all dipoles point
along $\vec n$, while for an isotropic structure $\langle \cos^2 \beta
\rangle = \frac{1}{2}$.  Thus, we define an alignment order parameter $p$
as
\begin{equation}
\label{eq:orderparameter}
p=2\langle \cos^2\beta \rangle-1,
\end{equation}
which is non-zero only for globally aligned structures and zero
otherwise. From our Monte Carlo simulations we compute $p$ as a
function of $T^{\star}$ and Poisson ratio for both lattice types.
As shown in \fig{fig1}a, on the square lattice a globally aligned
structure $p\rightarrow 1$ always develops regardless of Poisson ratio
when the noise level becomes sufficiently low. The transition
temperature $T^{\star}_c$ from a state without into a state with
long-ranged orientation correlations, i.e. where $p \neq 0$ for the
first time, decreases approximately linearly with $\nu$. This might be
attributed to the general enhancement of elastic interactions with
increasing Poisson ratio $W \sim (1+\nu)$. In
\fig{fig1}b we show how the probability distribution to
find a dipole with a given orientation $\beta$ evolves with increasing
noise levels. At low $T^{\star}$ almost all dipoles point along $\vec
n$ and the orientation distribution is strongly peaked; with
increasing $T^{\star}$ the distribution broadens and finally for
$T^{\star}>T^{\star}_c$ all orientations are - on a global scale -
equally likely (circle).  In contrast to the square lattice, where
long-ranged order always implies alignment, the type of long-ranged order
developed on a hexagonal lattice depends on Poisson ratio.
\fig{fig1}c shows that an aligned structure develops only
for $\nu<\nu_c\approx 0.32$.  Interestingly, in this case the relation
between the transition temperature $T^{\star}_c$ and Poisson ratio is
also reversed compared to the square lattice: $T^{\star}_c$ stays
roughly constant for $\nu<0.2$ and then drops quickly around
$\nu \approx \nu_c$. Moreover, towards $\nu_c$ the transition seems to
become more discontinuous and in our simulations $p$ suddenly jumps
from a low to a large value, indicating bistability and a first order
phase transition. The $T^{\star}_c$ scaling with $\nu$ might be
explained by simultaneous overall decrease in elastic signal strength
with decreasing $\nu$ within strings compensated by an increase in
lateral string-string coupling favoring global alignment at small
$\nu$.  For larger Poisson ratio ($\nu \ge 0.4$) we find $p=0$ at all
temperatures. At low $T^{\star}<T^{\star}_c$ a macroscopic isotropic
long-ranged ring-like order exists build up by small rings composed of
four dipoles oriented at 90 degrees with respect to each other. That
is, at $T^{\star}<T_c$ there are two equally large subpopulations of
dipoles oriented at 90 degrees with respect to each other. This
corresponds to the two peaks in the radial orientation distribution
shown in \fig{fig1}d $(\nu=0.5)$.  For $\nu=0.5$ the
transition temperature into the ordered ring-like structure
$T^{\star}_c$ is slightly larger than $1$ and significantly lower than
$T^{\star}_c \approx 2$ observed for the string structures on the
square lattice at equal dipole densities.

\begin{figure}
\begin{center}
\includegraphics[width=\columnwidth]{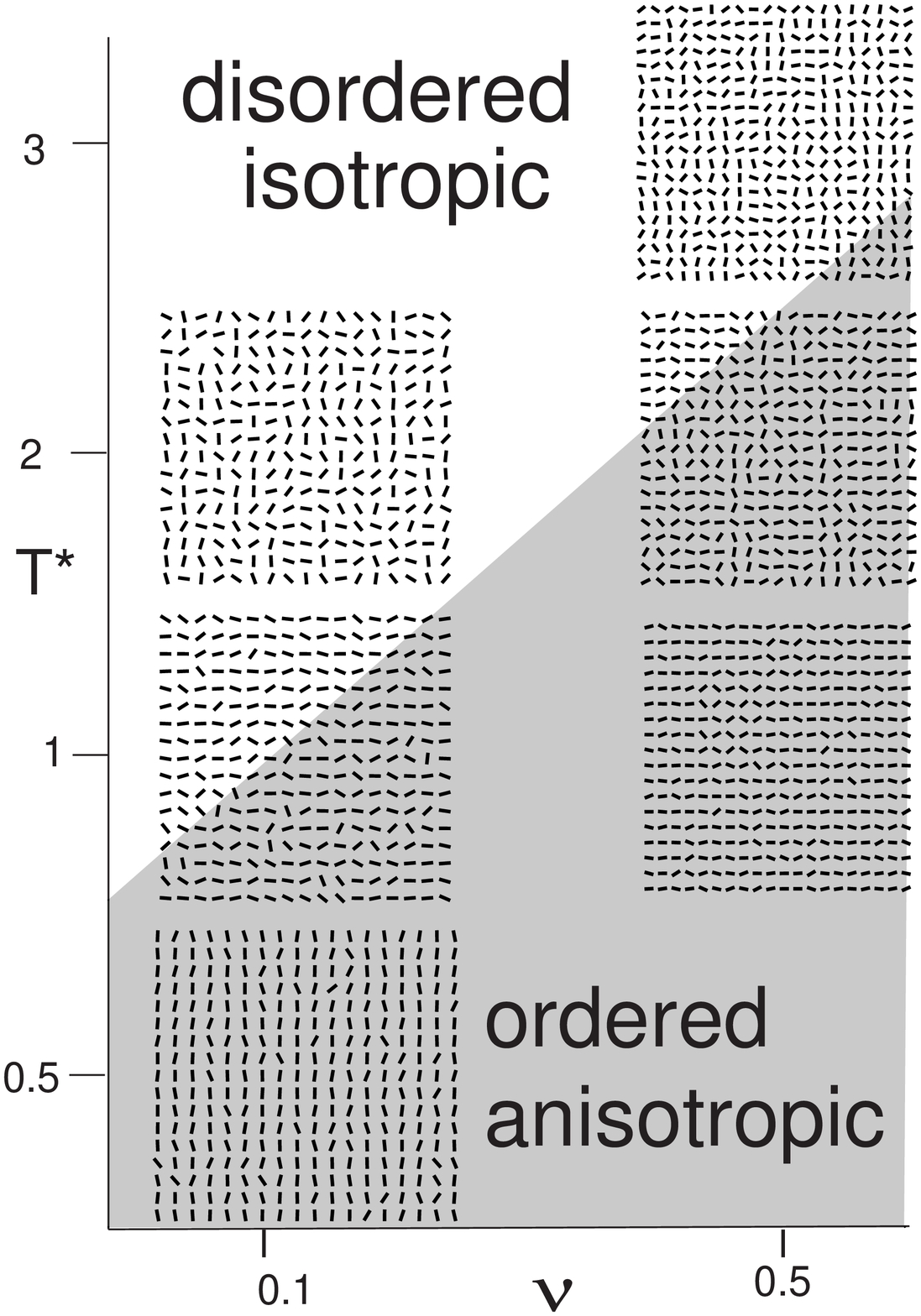}
\end{center}
\caption{Bischofs and Schwarz}
\label{fig:ordered_square}
\end{figure}

\begin{figure}
\begin{center}
\includegraphics[width=\columnwidth]{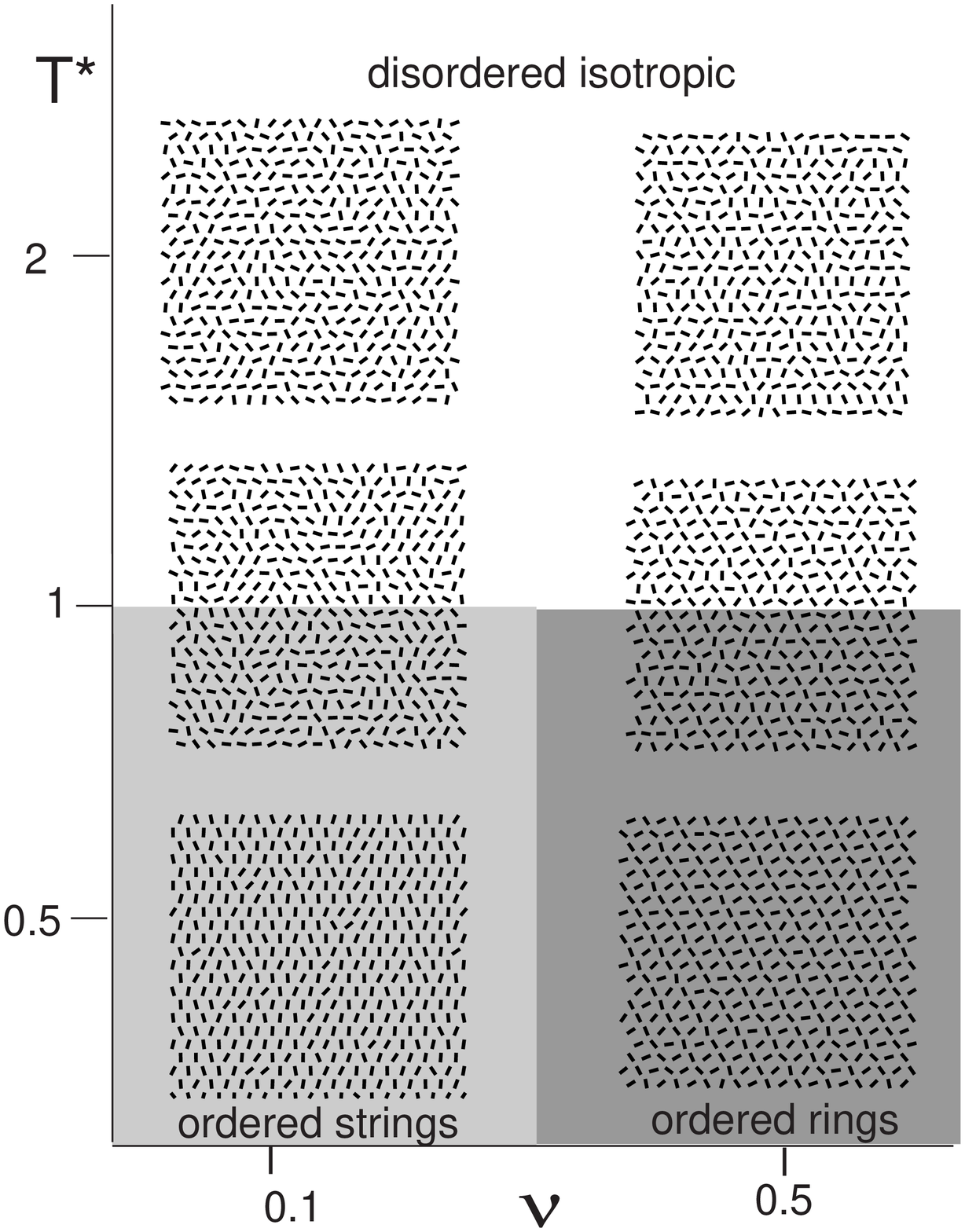}
\end{center}
\caption{Bischofs and Schwarz}
\label{fig:ordered_hex}
\end{figure}

In \ref{fig:ordered_square} and \ref{fig:ordered_hex} we summarize our
results for ordered structures in the form of $\nu-T^{\star}$ phase
diagrams for square and hexagonal lattices, respectively. We have
included representative snapshots of typical dipole configurations at
noise levels well below $T^{\star}_c$ in the globally ordered state,
approximately at the phase transition $T^{\star}_c$ and well above
$T^{\star}_c$ in the disordered phase for $\nu=0.1$ and $\nu=0.5$,
respectively.  It is important to note that despite the loss of
\textit{global} order at elevated noise levels
$T^{\star}>T_c^{\star}$, elastic interactions still \textit{locally}
manifest themselves in characteristic domain formation. For example,
at $T^{\star}=3$ elastic interactions still cause local order on a
square lattice and one finds short string-like domains preferentially
oriented along the principal lattice vectors.  On incompressible
substrates ($\nu = 0.5$) we furthermore often observe cooperative
excitations and the formation of ring-like domains.  This may reflect the
competition between ring- and string-like domains which is modulated
by the Poisson ratio. In fact, as we have shown before, at $\nu=0.5$
ring-like structures are only slightly disfavoured with respect to the
string-like structures \cite{uss:bisc05a}.  Nevertheless, it may seem
surprising that ring-like domains are excited so easily already at
very low $T^{\star}$ even before coexistence of the equally favorable
(1,0) and (0,1) domains occurs (see snapshot at $T^{\star}=1$). The
reason is probably a large domain interface penalty for two (1,0) and
(0,1) string domains compared to the interface penalty of a string and
an adjacent ring-domain.

For the hexagonal lattice local structural characteristics of elastic
interactions under noise are again string-like domains preferentially
oriented along directions specified by lattice geometry. For large
$\nu$, a characteristic signature for elastic interactions is the
formation of 4-cell-ring or ladder-domains with cells having
approximately perpendicular orientations with respect to each other.
Thus, although experimentally noise might be too strong to allow for
global ordering, elastic interactions might be still detected by their
local signatures of characteristic domain formation.

\subsubsection{Effect of weak positional disorder}

Experimentally cells may not adopt perfect lattice positions, in
particular when adhesive islands are used which are large compared to
cell size in order to minimize the effects of island shape on cellular
force distribution \cite{Parker02,Brock03}.  We therefore investigate
the effect of positional fluctuations at low thermal noise
intensities.  For this purpose we randomly displace the dipole
positions within a circle of radius $r$ around the lattice positions
before relaxing the orientations.  The degree of positional disorder
can be quantified by the ratio $f=r/b$ of the radius with respect to
the lattice constant $b$.  In the following we focus on the square
lattice and $\nu=0.5$, since synthetic polymer gels typically have
$\nu\approx0.5$. For this value of the Poisson ratio, \fig{fig1}
predicts a lattice geometry dependent transition from a string-like
structure on a square lattice to a ring-like structure on the
hexagonal lattice. However, rings are thermally excited easily on the
square lattice. Hence the string-like structure might also be easily
destabilized under positional disorder.

\begin{figure}
\begin{center}
\includegraphics[width=1\columnwidth]{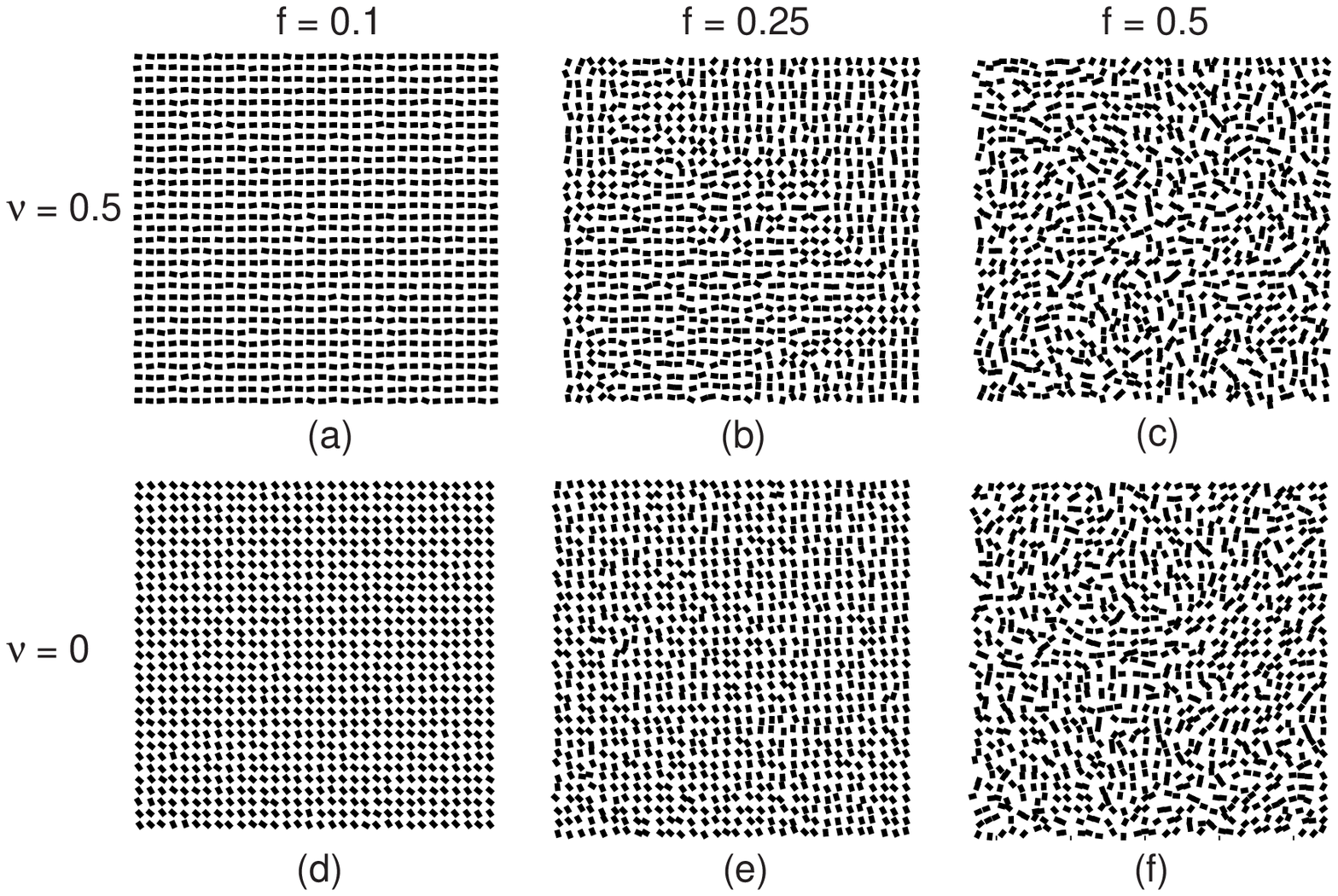}
\end{center}
\caption{Bischofs and Schwarz}
\label{fig:displaced}
\end{figure}

In \fig{fig:displaced} we show Monte Carlo snapshots at a
low reduced temperature ($T^{\star}=0.1$) for $f=0.1, 0.25, 0.5$. The
simulations show that a 10$\%$ deviation from the lattice positions
has only a minor effect on orientational ordering.  For $f=0.25$,
string-like as well as ring-like domains form and long-ranged
orientational order disappears. As for thermal noise, ring-domains
typically localize to interfaces between string-like domains with
perpendicular orientations. With increasing $f$ domains shrink and
structures appear increasingly disordered. Thus, moderate deviations
from the square lattice positions are not sufficient to
destabilize string-like structures with respect to ring-like
structures on incompressible substrates. Possibly because under
uniform positional noise dipoles still find themselves in a local
fourfold coordination with other dipoles (that is, each dipole has
typically for next neighbors as on the square lattice), string domains
are stabilized. We conclude that positional disorder affects structure
formation in a similar way as increasing the temperature $T^{\star}$
on a perfect lattice.

\subsection{Phase behavior on homogeneous substrates}

On a homogeneous substrate, non-motile cells usually adhere more or
less at random positions.  Therefore we next study typical structures
on elastic substrates with completely randomized but fixed positions
as described above. Cellular structure formation on elastic substrates
is now governed mainly by three control parameters, namely reduced
temperature $T^{\star}$, Poisson ratio $\nu$ and cell density. Since
in our model each cell is characterized by an exclusion radius $a$, we
introduce a reduced density
\begin{equation}
\rho^{\star}= \frac{N\pi a^2}{L^2},
\end{equation}
which is a dimensionless variable describing the ratio of the area
occupied by $N$ circular disks of radius $a$ to the area of the
(simulation) box with length $L$. 

\begin{figure}
\begin{center}
\includegraphics[width=1\columnwidth]{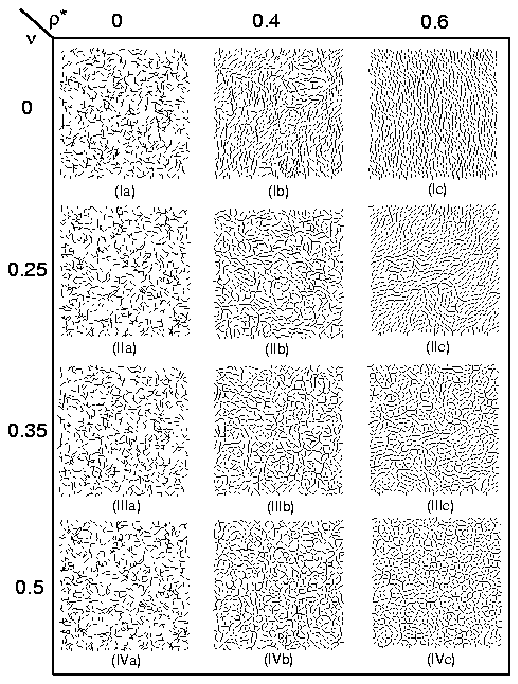}
\end{center}
\caption{Bischofs and Schwarz}
\label{fig:StaticNematic}
\end{figure}

\fig{fig:StaticNematic} shows typical snapshots of structures at
$T^{\star}=0.1$ for cells on an elastic substrate with $\nu=0, 0.25,
0.35, 0.5$ (top--bottom) for $\rho^{\star}=0, 0.4, 0.5$ (left--right).
At low densities cells predominantly optimize locally the interaction
between them by forming short strings with no long-range
correlation between string-like clusters. We attribute this finding to
the strong tendency of dipoles to form strings and the strong elastic
screening of string-string interactions \cite{uss:bisc05a}.  This
results in rather robust pattern formation that does not differ
qualitatively as the Poisson ratio is varied, see
\fig{fig:StaticNematic}a.  One expects that these patterns represent
typical cellular structures formed by strongly interacting cells when
cells in dilute concentrations are suspended on an elastic substrate
and adhere at random positions ($\rho^{\star}\rightarrow0$). With
increasing $\rho^{\star}$ the respective structures at low noise
intensity show a strong dependence on the Poisson ratio $\nu$. For
incompressible substrates ($\nu = 0.5$) we find isotropic ring-like
structures, see \fig{fig:StaticNematic}(IVc).  With decreasing Poisson
ratio string-like patterns emerge. For $\nu=0.35$ we find coexistence
of string-like and ring-like domains, compare
\fig{fig:StaticNematic}(IIIc). For substrates with Poisson ratio below
$\nu<0.32$ string-like structures dominate at intermediate densities.
With increasing density strings start to interact and domains of
aligned parallel strings form which increase in size with increasing
$\rho^{\star}$, see \fig{fig:StaticNematic}(IIb,IIc). Highly
compressible substrates ($\nu \to 0$) favor cell alignment along a
common direction at high cell densities and in
\fig{fig:StaticNematic}(Ic) we find a globally aligned structure.
Here, the rotational symmetry of the structure is spontaneously broken
along an arbitrary direction in space despite lacking long-ranged
positional order. In this respect we may speak of a nematic structure
in analogy to nematic phases formed by liquid crystals which are
characterized by orientational order but positional disorder.

\begin{figure}
\begin{center}
\includegraphics[width=\columnwidth]{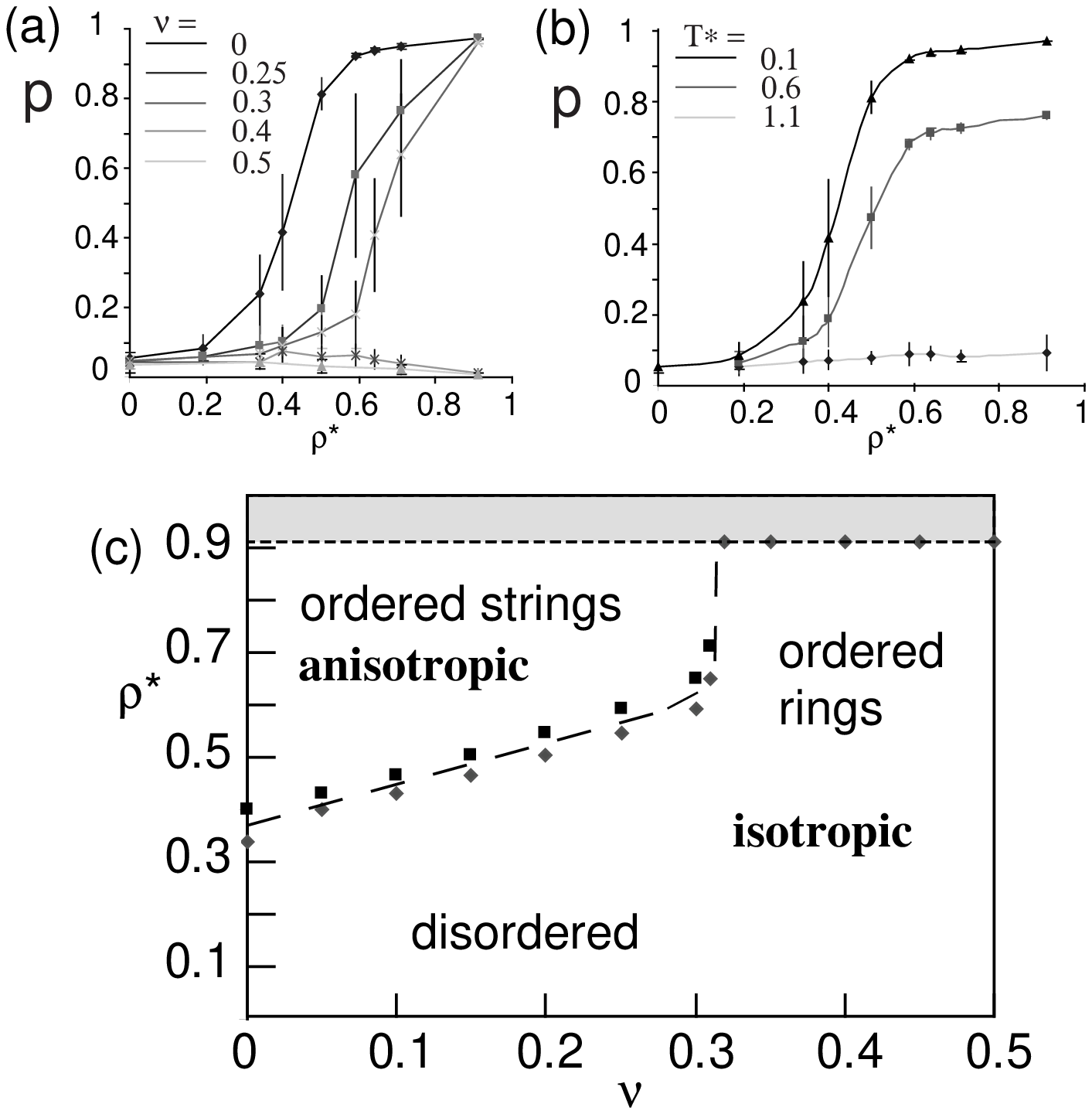}
\end{center}
\caption{Bischofs and Schwarz}
\label{fig:nematicsummary}
\end{figure}

As before one can quantify the development of anisosotropic
long-ranged order with the order parameter $p$ defined in
\eq{eq:orderparameter}. In contrast to the lattice structures where
preferred string direction $\vec n$ was determined by lattice
geometry, the orientation of the director $\vec n$ on homogeneous
substrates is arbitrary. This is taken into account by defining a
two-dimensional analog of the nematic order parameter $p$ used in the
theory of liquid crystals \cite{deGennes95}. We first introduce the
ordering matrix
%\begin{equation}
$Q_{ij}=\frac{1}{N}\sum_{\alpha=1}^N
(l_i^{\alpha} l_j^{\alpha} - \frac{1}{2}\delta_{ij})$,
%\end{equation}
where $\vec l^{\alpha}$ is the orientation vector of the $\alpha$`th particle
and the sum runs over all particles in the simulation box. The
largest eigenvalue $\lambda$ of the symmetric ordering
matrix $Q$ corresponds then to the order parameter $ p = 2 \lambda$. As before
$p$ measures the degree of orientational order with respect to the current
director $\vec n$, which is the corresponding eigenvector to the
maximal eigenvalue. $p=1$ implies nematic and $p=0$ isotropic structures.
Since we freeze in positions, we first thermally average $p$ for a fixed
configuration of the dipole positions and subsequently average
over at least 20 random position configurations obtained for the
same $\rho^{\star}$. 
In \fig{fig:nematicsummary} we show the quantitative effects of the
reduced density $\rho^{\star}$, Poisson ratio $\nu$ and reduced
temperature $T^{\star}$ on nematic ordering. At constant temperature
$T^{\star}=0.1$ a nematic structure forms beyond a critical density
$\rho^{\star}_c(\nu)$ and below a critical value of the Poisson ratio,
see \fig{fig:nematicsummary}a. The degree of structural alignment
increases with increasing $\rho^{\star}$ and approaches $p\rightarrow
1$ toward the maximal $\rho^{\star}=0.907$, which is the maximal
packing density achieved for disks arranged on a hexagonal lattice.
For $\nu \ge 0.4$ no nematic ordering $p=0$ occurs at any density
$\rho^{\star}$ or temperature $T^{\star}$, because ordered structures
are ring-like.  In addition to the density-dependent ordering
transition, the nematic structure is also destabilized by increased
noise. In \fig{fig:nematicsummary}b we plot $p$ for $\nu=0$ at
different values of the reduced temperature $T^{\star}$. The critical
density $\rho^{\star}_c$ to obtain a nematic structure increases with
$T^{\star}$ and $p = 0$ for any any $\rho^{\star}$ when $T^{\star}$
exceeds a critical $T^{\star}_c\approx 1$, which corresponds
approximately to $T^{\star}_c$ measured on the hexagonal lattice.

In \fig{fig:nematicsummary}c we show the phase diagram in the
$\nu$-$\rho^{\star}$-plane for $T^{\star}=0.1$. Diamonds mark $p
<0.4$, squares $p >0.4$ and the dashed line denotes our estimate for
the isoline $p =0.4$. We find three different phases: a high-density
anisotropic string-like structure, a high-density isotropic ring-like
structure and a low density disordered structure.  The general
features of this phase diagram are very similar to the thermal phase
diagram for hexagonal lattices shown in \fig{fig:ordered_hex} but with
the role of $T^{\star}$ and (inverse) density $\rho^{\star}$
exchanged.  Note, that at $\rho^{\star}=0.907$ dipoles in fact adopt
hexagonal lattice positions, because disks must order into a hexagonal
lattice due to the non-overlap constraint. Thus, reducing the density
from $\rho^{\star}=0.907$ in our model is in some sense equivalent to
introducing position fluctuations around hexagonal lattice positions
(see above), which qualitatively has similar consequences as raising
the effective temperature.  These results also show the importance of
local position geometry and dipole coordination. At high
$\rho^{\star}$ dipoles have six next neighbors and all interactions
are equal in magnitude.  Because elastic interactions are strongly
anisotropic and dependent on Poisson ratio, non-trivial structure
formation as a function of local dipole coordination and Poisson ratio
results. Decreasing $\rho^{\star}$ the number of next neighbors
decreases and at low $\rho^{\star}$ dipoles typically only have one or
two next neighbors, which favors dipole alignment and string
formation.  Due to screening of the traction patterns in a string,
adjacent strings hardly interact and we obtain a largely disordered
structure even at low $T^{\star}$.  In our calculations the degree of
position correlation is controlled by $\rho^{\star}$, which therefore
plays a similar role as lattice geometry before. In contrast to
relative density effects described by $\rho^{\star}$, absolute density
effects can be subsumed into the effective temperature $T^{\star}$,
see \eq{eq:Tstar}.  This is due to the fact that by
\textit{increasing} cell density the screening efficiency of
string-string interactions also \textit{increases}. The only way to
overcome screening and introduce lateral coupling between dipoles is
to insure that dipole-dipole distances within a string and the
distance to an adjacent string are comparable. Increased dipole
coordination, however, is controlled by $\rho^{\star}$ rather than by
$\rho$.

We note that the order-disorder transition with $\rho^{\star}$ might
also represent a \textit{spin-glass transition}. Such transitions are
common for magnets, where spins represent magnetic moments. In magnets
global ordering of spins can result in a ferromagnetic state, where
all spins point along a common direction and a macroscopic magnetic
field builds up. Anti-ferromagnetic order results when spins are
globally ordered but with equal numbers of up and down spins, such
that no macroscopic polarization builds up. A spin-glass transition
occurs when e.g.\ due to disordered positions of spins no ferro- or
anti-ferromagnetic order develops even at low temperatures. In
analogy, dipoles in our model represent spins. The aligned structure
therefore might be called a \textit{ferroelastic phase}, because a
macroscopically anisotropic polarization stress develops
spontaneously. In analogy, the ring-like structure might be called an
\textit{antiferroelastic phase}, because the macroscopic stress
exerted by the dipoles remains isotropic.  When $\rho^{\star}<\rho_c$,
no global order develops even in the limit $T^{\star}\rightarrow 0$;
instead, many equally favorable states exist. Because we do not allow
dipoles to adjust positions, they cannot relax into an ordered
ferro- or anti-ferroelastic state. Thus, the disordered low
temperature phase might also represent a spin glass. There is an
interesting difference between thermally disordered states and density
disordered states at low T. In both cases, dipoles can be aligned by
application of an external field. The former might be called a
\textit{paraelastic phase} because it relaxes into the unpolarized
disordered state after the external field is switched off with a
single time-scale $\tau \propto (T-T_c)^{\gamma}$. In contrast, the
later might represent a spin-glass state and relaxation into the
disordered state cannot be described by a single power law. In fact a
characteristic feature of spin glasses is a fast relaxation to a
remanent polarization and then a slow relaxation into the unpolarized
state, which may exceed experimental timescales.

\section{Elastic effect in 3D environments}

\begin{figure}
\begin{center}
\includegraphics[width=0.7\columnwidth]{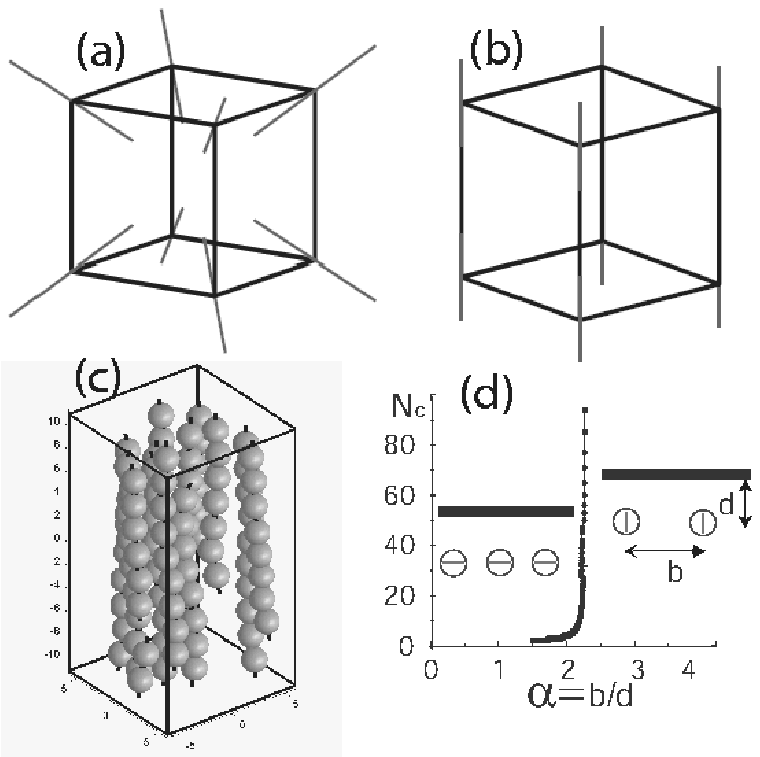}
\end{center}
\caption{Bischofs and Schwarz}
\label{fig:3DSummary}
\end{figure}

Although cell behaviour in 3D is not the focus of this paper, we now
spell out a few predictions resulting from our model in this case.
Calculations for cells positioned on a simple cubic lattice suggest
that in incompressible environments ($\nu = 1/2$) the optimal
structure is effectively isotropic with a hedgehog-like unit cell,
where all dipoles at the corners point to the cube's center, see
\fig{fig:3DSummary}a, while for $\nu=0$ spontaneous symmetry breaking
along a principal lattice lattice vector occurs, see
\fig{fig:3DSummary}b. This indicates that similar transitions between
ring-like and string-like structures as a function of Poisson ratio
$\nu$ exist in 3D as they do in 2D. String-like structures might also
be favored by anisotropies in the effective mechanical properties of
the environment. For example, fiber alignment or external strain
fields might cause cell alignment with the external perturbation,
which could be further stabilized by elastic interactions between
cells. In \fig{fig:3DSummary}c we show a corresponding snapshot of a
Monte Carlo simulation with $100$ cells modeled as hard spheres with
an elastic dipole moment at their center.  Here we allowed for both
orientational and positional degrees of freedom ($T^{\star}=2, \nu =
1/2$), where the position dependence of $W$ from \eq{eq:1Dstacka} is
used for additional Monte Carlo moves.  Moreover we applied a
homogeneous strain field along the $z$-direction. Cells respond by an
alignment with the external field and the formation of strings as has
been observed experimentally \cite{Eastwood98}. Without external field
but with mobile cells we typically observe formation of "networks" of
cells at low temperature and intermediate densities and disconnected
uncorrelated strings at elevated noise levels and low densities,
respectively. In general, we expect that the low to intermediate
density regime of a three 3D phase diagram for mobile elastically
interacting force dipoles will resemble the phase behavior of electric
dipolar fluids \cite{Tlusty,Weis}. In both systems string formation
dominates and string-string interactions are screened in the same way
\cite{Allen}. At high dipole densities our lattice calculations in 3D
suggest that phase transitions as a function of Poisson ratio may
occur and we find evidence that again a low Poisson ratio of the
material tends to favor global cell alignment.

In finite 3D environments the geometry and boundary condition of the
sample may also affect structure formation. The presence of a boundary
modifies the direct elastic interaction between cells by boundary
induced strain fields, which, depending on boundary condition,
introduce either attractive or repulsive contributions to the elastic
interaction plus a direct interaction term with the boundary
\cite{Bischofs03,Bischofs04}.  An instructive example for the
competition between the direct interaction of a cell with the boundary
and cellular interactions is the elastic half space with a clamped
boundary. When cells are located close to the boundary, the direct
interaction with a clamped boundary favors cellular orientations
pointing toward the surface \cite{Bischofs03,Bischofs04}. On the other
hand, interactions between cells favor the formation of strings and
thus parallel orientations. Our calculations suggest that the
transition between these two configurations is a function of the ratio
$\alpha=b/d$, where $d$ is the distance to the boundary and $b$ is the
distance between cells, and the number of interacting cells $N$. Above
a critical value $\alpha_c \approx 2.3$ the direct interaction with
the boundary dominates and cells are expected to point toward the
surface. For $\alpha<\alpha_c$ orientations in parallel to the clamped
surface are favored when sufficiently many cells are present, $N >
N_c(\alpha)$. In \fig{fig:3DSummary}d we plot $N_c$ as a function of
$\alpha$. Thus, collective effects may alter preferred cell
organization close to clamped boundaries. In contrast for a free
surface the direct interaction with boundary favors parallel
alignment, which is further stabilized by collective elastic effects
between cells. This may explain the robust parallel alignment of cells
with respect to free surfaces which has been observed
experimentally \cite{Bell79,Takakuda96}.

\section{Conclusion and Outlook}

Using a simple mathematical model and extensive Monte Carlo
simulations, we have analyzed structure formation due to elastic
effects for fibroblast-like cells with anisotropic force patterns. Our
model predicts in a quantitative way how structure formation is
controlled on soft elastic substrates by the Poisson ratio $\nu$, the
reduced cell density $\rho^{\star}$, the relative strength of elastic
interactions with respect to noise (represented by the reduced
temperature $T^{\star}$) and the geometry of cell positioning. Up to
intermediate cell densities, we predict the formation of uncorrelated
strings of cells (paraelastic phase). This finding can be explained by
noting that the cellular traction patterns in strings screen each
other, as we have demonstrated before by an analytical calculation
\cite{uss:bisc05a}.  For high cell densities, we find an interesting
competition between isotropic ring-like and anisotropic string-like
structures. The string-like (ferroelastic) phase is reminiscent of the
nematic phase for liquid crystals. In contrast, the nematic order
parameter vanishes in the ring-like (anti-ferroelastic) phase, because
local ordering principles preclude nematic order on the macroscopic
scale.

The Poisson ratio $\nu$ allows to switch between these phases. It
strongly affects cellular structure formation due to the non-trivial
way in which stress and strain propagates in the elastic environment.
The antiferroelastic phase is expected to occur on
synthetic elastic substrates made from e.g.\ polyacrylamide or
polydimethylsiloxane, which are characterized by $\nu \approx 1/2$. In
order to obtain the nematic phase, new kinds of polymer gels for cell
culture are needed, with considerable lower values for the Poisson
ratio (namely below $\nu_c=0.32$). One possible route might be the
design of biocompatible polymer gels with large meshsizes but little
hydration, thus avoiding the incompressibility effect of bound water.
As it is an open issue on which time and length scales cells sense the
mechanical properties of their environment \cite{c:disc05}, design and
testing of new materials is crucial for progress in this field.

Like the Poisson ratio, also the other determinants of cellular
structure formation are in principle accessible experimentally. The
Young modulus $E$ directly affects the reduced temperature and can
easily be varied in experiments. Nevertheless it does not appear to be
a reasonable control variable for experiments, mainly because the
predicted structure formation might be expected to take place only in
a small range of $E$-values, namely the ones corresponding to
physiological rigidities (around kPa). Other control parameters of
interest are cellular contractility (which affects $T^{\star}$ through
the force dipole moment $P$ and which might be varied for example by
administering LPA, which stimulates Rho-mediated contractility) and
the average distance $b$ between cells (although, in order for the
force dipole approximation to remain valid, the typical distance
between cells should not be smaller than the typical cell length). The
most important experimental control parameter next to the Poisson
ratio $\nu$ might be the geometry of cell positioning, which can be
controlled with patterning techniques (including microcontact printing
in two dimensions).

Elastic effects are likely to contribute to large scale tissue
organization and to the structure-function relationship of tissues.
Our results indicate that for sparsely populated tissues, elastic
interactions, despite their long-ranged character, might have only a
rather local effect. This is a result of the local tendency for cells
to align into strings combined with the resulting screening of elastic
fields. In this regime the system remains macroscopically disordered
over a wide range of variations in material properties and cell
density, effectively making the composite material of matrix and cells
robust against perturbations. It is well known that the sparsely
populated connective tissue is a rather disordered structure. Our
simulations suggest that without macroscopic external fields elastic
interactions between cells favor isotropic disordered structures. In
this case the average stress in the tissue resulting from cells is
isotropic, that is $\sigma_{ij}=P \rho \delta_{ij}$, where $\rho$ is
the cell density.  The situation changes when an external field is
applied (e.g.\ resulting from a wound). Then highly aligned structures
may prevail. In this case cells orient their mechanical activity in
such a way that the average cellular stress in the medium
$\sigma_{ij}=\rho \langle P_{ij} \rangle$ becomes anisotropic in
response to homogeneous uniaxial stress, that is, the cellular stress
is directed opposite to the externally applied stress (e.g. as to
close the wound). It is an interesting question what happens after the
external alignment field has decreased to zero. Our calculations
suggest that elastic interactions may stabilize an aligned structure
even in the absence of external fields. In particular, disordered
tissues (like the connective tissue) may become aligned due to an
external field. Due to elastic interactions the transition from the
aligned state back into the normal disordered state upon switching off
the external field could be dramatically slowed down, in analogy to
similar effects observed in magnetic spin glasses.  This could have
implications for wound healing and scar formation.  Polarization also
occurs spontaneously in our model. Interestingly, this state reminds
of the disease state called \textit{contracture}, when certain tissues
(like skin or muscle) start to tighten up, eventually preventing
movement of body parts. It is an interesting speculation whether the
mechanical properties of extracellular matrices are altered by
(diseased) cells in such a way as to effectively induce variations in
the Poisson ratio, which could trigger transitions between ferro- and
anti-ferroelastic states. In order to investigate this point in more
detail, our model should be extended to include viscoelastic and
plastic effects as well as fiber degrees of freedom.

Mechanotaxis also plays an important role for \textit{dynamic}
rearrangement of groups of cells in tissues, including development,
wound healing, capillary sprouting and tumor growth.  The work
presented here builds on experimental observations which suggest that
cells migrate toward high strain areas and orient their mechanical
activity in such a way as to pull back in response to external tensile
strain.  Such tensile strain is certainly present e.g.\ close to
wounded areas or tumors. Future studies will show in which sense our
model can be adapted to these more specific situations. The current
study shows in a general way that many interesting structure formation
processes arise from mechanical effects in communities of
mechanosensitive cells and therefore might help not only to understand
the corresponding physiological situations, but also to program new
tissue functions by guiding the design of novel material with
appropriate mechanical properties.

\section*{Acknowledgments}

We thank Phil Allen, Sam Safran, Assi Zemel and Thomas Pfeifer for
helpful discussions. This work was supported by the German Research
Foundation (DFG) through the Emmy Noether Program as well as by the
Center for Modelling and Simulation in the Biosciences (BIOMS) at
Heidelberg.

\newpage

%\bibliography{BPJ,uss,b,c}
%\bibliographystyle{unsrt}

\newpage
\begin{center}
\textbf{\Large Figure Captions}
\end{center}

Figure~1: Effect of noise on global alignment order. (a) The order
parameter $p$ for alignment on a square lattice as a function of noise
$T^{\star}$ for Poisson ratio $\nu=0,0.1,0.2,0.3,0.4,0.5$ (from left
to right). For low noise, aligned structures develop for all values of
the Poisson ratio. (b) The radial distribution of the dipolar
orientations on a square lattice with $T^{\star}=1.5,2,3$ for Poisson
ratio $\nu = 1/2$. The two peaks correspond to an aligned structure
with bipolar symmetry below $T_C^{\star} = 3$. (c) On a hexagonal
lattice, $p = 0$ for $\nu=0.4,0.5$ and $p \rightarrow 1$ for
$\nu=0.3,0.2,0.1,0$ from left to right. At low noise, aligned
structures are unstable for large values of the Poisson ratio. (d) The
radial distribution function on a hexagonal lattice with
$T^{\star}=0.5,1,2$ for Poisson ratio $\nu = 1/2$. The four peaks
correspond to a ring-like structure below $T_c^{\star}=2$.
\\
\\
Figure~2: Schematic phase diagram as a function of noise level $T^*$
and Poisson ratio $\nu$ for the square lattice. For illustration,
typical Monte Carlo snapshots are added. Both decreasing $T^*$ or
increasing $\nu$ favors the ordered string-like phase. For $\nu=0.1$,
the dipoles typically weakly fluctuate around the string-like ground
state. For $\nu=0.5$, the dominant fluctuations around the optimal
state are ring-like.
\\
\\
Figure~3: Schematic phase diagram as a function of noise level $T^*$
and Poisson ratio $\nu$ for the hexagonal lattice. For illustration,
typical Monte Carlo snapshots are added. Ordered phases are favored
by decreasing $T^*$, but the ordered phases are string- and ring-like
for small and large values of $\nu$, respectively.  
\\
\\
Figure~4: Snapshots of Monte Carlo simulations at reduced temperature
$T^{\star}=0.1$ for cellular force dipoles randomly displaced from an
ideal square lattice for $f=0.1,0.25,0.5$ from left to right for
$\nu=0.5$ (a-c) and $\nu=0$ (d-f), respectively. Increased positional
disorder $f$ destroys long-ranged orientational order in a similar way
as an increase in $T^{\star}$ on a perfect lattice.
\\
\\
Figure~5: Snapshots of Monte Carlo simulations at reduced temperature
$T^{\star}=0.1$ for $N=1024$ cellular force dipoles at random, but
fixed positions on an elastic substrate with $\nu=0,0.25,0.35,0.5$,
respectively (top-bottom). The reduced area density increases from
left to right as $\rho^{\star}=0, 0.4, 0.6$, respectively, while the
average cell density $\langle \rho \rangle$ remains constant.
\\
\\
Figure~6: Averaged nematic order parameter $p$ as a function of model
parameters. (a) Simulation results for $\nu=0, 0.25, 0.3$ (left-right)
and $\nu=0.4,0.5$ (bottom), respectively. Thus nematic ordering occurs
only on compressible substrates and above a critical density. (b)
Simulation results for $\nu=0$ at $T^{\star}=0.1, 0.6, 1.1$
(top-bottom). Thus nematic ordering disappears above a certain noise
level. (c) Phase diagram for $T^{\star}=0.1$ for dipoles on elastic
substrates obtained by Monte Carlo simulations. All points below
diamonds have $p<0.4$ and all points above squares yield $p>0.4$. The
long dashed line is our estimate for the isoline with $p=0.4$. The
horizontal dashed line marks the maximal $\rho^{\star}$ possible for a
hexagonal lattice.
\\
\\
Figure~7: Collective effects in 3D elastic media. (a,b) The optimal
structure in a cubic lattice depends on the Poisson ratio $\nu$. In
incompressible media ($\nu = 1/2$) the isotropic hedgehog structure is
most favorable (a) while in highly compressible media ($\nu = 0$)
elastic interactions favor aligned structures (b). (c) Cells in
external strain fields form strings running along the direction of
stretch due to interactions with external strain and elastic cell-cell
interactions. (d) Collective effects can modify preferred cell
organization close to a clamped boundary, which for single cells is
perpendicular. For small intercellular distances ($\alpha = b/d <
2.3$) and sufficiently many cells in a string ($N > N_c$), the
parallel orientation becomes more favorable.

\newpage


\end{document}